\documentclass[aps,prb,twocolumn,superscriptaddress,showpacs]{revtex4}
\usepackage{graphicx}
\usepackage{dcolumn}
\usepackage{amsmath}
\usepackage{epsfig}

\begin{document}

\newcommand{\etal}[0]{\textit{et al.}}
\newcommand{\alumina}[0]{Al$_2$O$_3$}
\newcommand{\Kalumina}[0]{$\kappa$-Al$_2$O$_3$}
\newcommand{\Aalumina}[0]{$\alpha$-Al$_2$O$_3$}
\newcommand{\Galumina}[0]{$\gamma$-Al$_2$O$_3$}
\newcommand{\bea}[0]{\begin{eqnarray}}
\newcommand{\eea}[0]{\end{eqnarray}}
\newcommand{\nn}[0]{\nonumber}
\newcommand{\mtext}[1]{\mbox{\tiny{#1}}}
\renewcommand{\vec}[1]{{\bf #1}}
\newcommand{\op}[1]{{\bf\hat{#1}}}
\newcommand{\galpha}[0]{$\alpha$}
\newcommand{\gbeta}[0]{$\beta$}
\newcommand{\ggamma}[0]{$\gamma$}
\newcommand{\gdelta}[0]{$\delta$}
\newcommand{\gepsilon}[0]{$\epsilon$}
\newcommand{\gphi}[0]{$\phi$}
\newcommand{\gvarphi}[0]{$\varphi$}
\newcommand{\geta}[0]{$\eta$}
\newcommand{\gtheta}[0]{$\theta$}
\newcommand{\gomega}[0]{$\omega$}
\newcommand{\gchi}[0]{$\chi$}
\newcommand{\gxi}[0]{$\xi$}
\newcommand{\gkappa}[0]{$\kappa$}
\newcommand{\del}[1]{\partial_{#1}}
\newcommand{\av}[1]{\langle #1\rangle}
\newcommand{\bra}[1]{\langle #1\right|}
\newcommand{\ket}[1]{\left| #1\right\rangle}
\newcommand{\bracket}[3]{\langle #1 | #2 | #3\rangle}
\newcommand{\sproduct}[2]{\langle #1 | #2\rangle}
\newcommand{\Bracket}[3]{\left\langle #1 \left| #2 \right| #3\right\rangle}
\newcommand{\sign}[1]{\mbox{sign}\left(#1\right)}
\newcommand{\mc}[3]{\multicolumn{#1}{#2}{#3}}
\newcommand\T{\rule{0pt}{2.6ex}}
\newcommand\B{\rule[-1.2ex]{0pt}{0pt}}

\title{
Stacking and band structure of van der Waals bonded graphane multilayers}

\author{Jochen Rohrer}
\affiliation{%
BioNano Systems Laboratory, 
Department of Microtechnology, 
MC2, 
Chalmers University of Technology, 
SE-412 96 Gothenburg
}%
\author{Per Hyldgaard}%
\email{hyldgaar@chalmers.se}
\affiliation{%
BioNano Systems Laboratory, 
Department of Microtechnology, 
MC2, 
Chalmers University of Technology,
SE-412 96 Gothenburg
}%

\date{\today}

\begin{abstract}
We use density functional theory 
and the  van der Waals density functional (vdW-DF) method 
to determine the binding separation in bilayer and bulk graphane
and study the changes in electronic band structure
that arise with the multilayer formation.
The calculated binding separation (distance between center-of-mass planes)
and binding energy are $4.5-5.0$~\AA\ ($4.5-4.8$~\AA) and $75-102$~meV/cell ($93-127$~meV/cell)
in the bilayer (bulk), depending on the choice of vdW-DF version.
We obtain the corresponding band diagrams using calculations
in the ordinary generalized gradient approximation
for the geometries specified by our vdW-DF results,
so probing the indirect effect of vdW forces on electron behavior.
We find significant band-gap modifications by up to -1.2~eV (+4.0~eV)
in various regions of the Brillouin zone,
produced by the bilayer (bulk) formation.
\end{abstract}


\maketitle

\section{Introduction}
Selective modification of band gaps (band-gap engineering)
by atomic-scale design of materials is a powerful concept
in electronic  and photonic  development.\cite{BGE}
Band gaps can be altered by, for example, 
introducing dopants, defects or by exploiting 
finite size effects.\cite{defects,finiteSize}
The physical origin of band-gap variations is
a modification in the charge distribution
in concert with wavefunction hybridization and modification.

Dispersive or van der Waals (vdW) interactions\cite{vdWDFreview} also 
alter the distribution of electronic charges
and hence the electron band-structure.
A \textit{direct effect} is evident, 
for example, by considering the formation of the
double-dipole configuration\cite{Thonhauser2007}
which is the electrostatic signature and inherent nature
of a pure vdW binding.\cite{vdWDFreview,Thonhauser2007}
In addition, there are also \textit{indirect, geometry-induced effects} of 
vdW binding on electron behavior. 
These indirect effects arise when two material fragments 
come close to one another, 
thereby changing the local electron environments as compared to isolated fragments.
For example, vdW-binding can cause smaller amounts of net charge 
transfer within individual vdW-bonded fragments.\cite{Londero2010}
Also, wavefunction hybridization will certainly arise 
when material fragments approach one another,
even if this hybridization does not significantly contribute
to the binding itself (in purely dispersive interaction).
Wavefunction hybridization and Pauli exclusion 
cause scattering of the surface-state electrons
in physiosorption of acenes and quinones on Cu$(111)$,
even if there is no net charge transfer.\cite{sliding, sliding2}

The strength of vdW interactions 
in surface/adatom systems (including adclusters or complete overlayers)
can be controlled by the precise choice of the surface material\cite{JacobssenGrapheneOnMetals, RohrerPhD, RohrerGrapheneOnSiC}
and by the surface morphology (flat, stepped, pyramidal, \textit{etc.}).\cite{KelkkanenBenxeneOnMetals}
It is therefore important to quantify the extent to which van der Waals (vdW)
bonding can modify the electron dispersion,
that is, the band structure.
Such a study is now possible, 
since recent development of the vdW  density functional (vdW-DF) method
\cite{Thonhauser2007,vdWDF,vdWDF2,C09}
enables systematic (theoretical) explorations 
of  bonding in sparse materials
within density functional theory (DFT).\cite{vdWDFreview}

\begin{figure}[b]
\begin{tabular}{c}
\includegraphics[width=8.4cm]{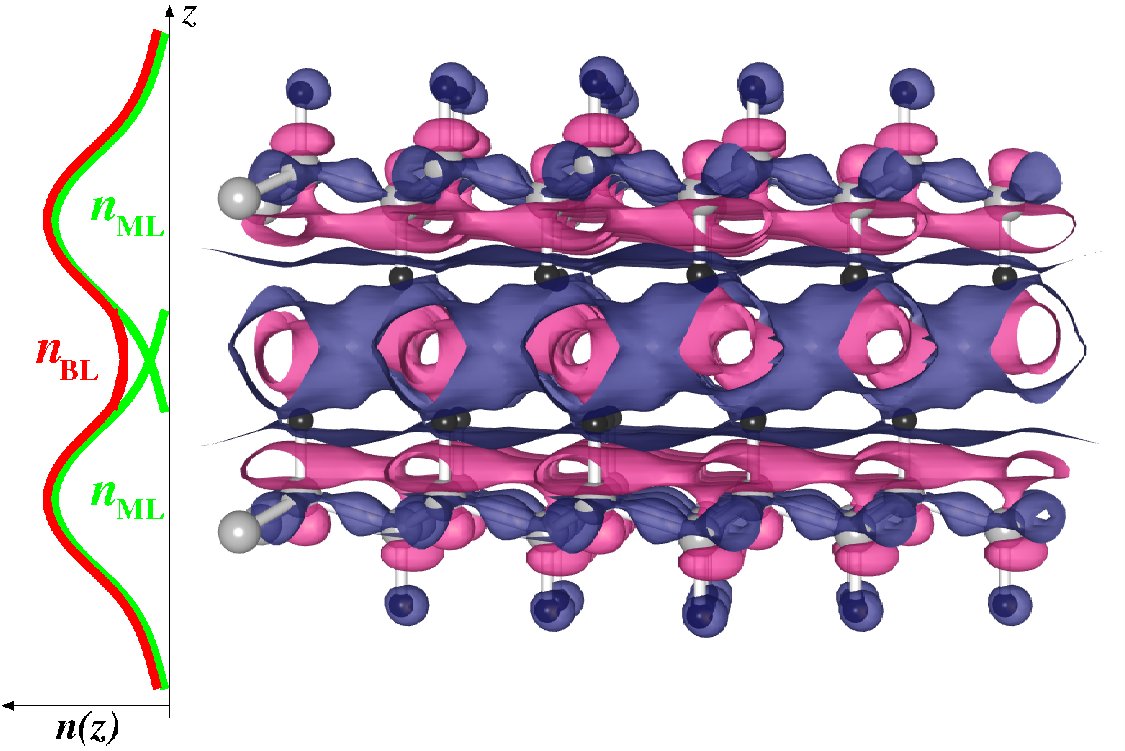}
\end{tabular}
\caption{
\label{fig:ChargeRearrangements}
(Color) Charge density in vdW-bonded graphane bilayers.
The left panel shows a schematics of the non-selfconsistent 
bilayer (BL) charge density obtained by adding two 
monolayer (ML) densities and resulting
into a charge enhancement in the interlayer region.
The right panel depicts calculated self-consistent 
(on the GGA level) charge rearrangements.
Our Bader analysis\cite{BaderAnalysis1,BaderAnalysis2} shows that
the charge rearrangements are \textit{not} 
accompanied by any significant net charge transfer
into the middle region.
Color coding: 
light large spheres represent C atoms,
black small spheres represent H atoms;
accumulation of charge is indicated by the pink 
(light) isosurface,
depletion by the purple (dark) isosurface.
}
\end{figure}

In this paper we investigate \textit{indirect}, geometry-induced effects 
of vdW bonding on electronic structure.
In particular we separately study the effects
of (1) the enhanced charge density
in  regions where the tails of vdW-bonded material fragments overlap,
see left panel of Fig.~\ref{fig:ChargeRearrangements},
(2)~the hybridization of wave functions,
and
(3)~the redistribution of charge density
due to mechanisms that are not inherent to the vdW interaction
(such as local displacements due to electrostatics and Pauli-repulsion),
see right panel of Fig.~\ref{fig:ChargeRearrangements}.

We focus on the band structure of layered systems of the
macromolecule graphane,\cite{ref:Sofo_Graphane_2007,ref:Elias_Graphane_2009}
a fully hydrogenated derivative of graphene.\cite{ref:Graphene}
The top panel of Fig.~\ref{fig:structures} shows the atomic structure
of (the stable chair conformation of) monolayer (ML) graphane,
consisting of a (slightly buckled) graphene backbone
with H atoms attached in alternating fashion above and below 
the carbon plane.
Bilayer graphane with possible high-symmetry structure shown 
in the bottom panels of Fig.~\ref{fig:structures}
present a possible new system.\cite{BLGraphane1,BLGraphane2}
In addition, we also include a study of a possible bulk-graphane crystal.

Graphane adds to the wealth of carbon-based materials
that are considered as promising materials
for near-future nanoelectronic devices.\cite{ref:Devices1,ref:Devices2}
Pure graphene has a zero band gap and extraordinary conduction properties.
Electronic devices, however, also require semiconducting and insulating materials.
Such materials can be obtained from pure graphene as derivatives
either in the form of graphene  nanoribbons\cite{GNRExp,GNRModel}
or by chemical modification through adsorbates.\cite{ChemicalModification,BGE_Eriksson}
Monolayer graphane belongs to the last-mentioned group of derivatives.
DFT calculations predict  a large band-gap semiconductor nature;\cite{ref:Sofo_Graphane_2007}
the more advanced GW method\cite{GW} predicts an insulating nature.\cite{ErikssonGW}
Graphane has been proposed theoretically to serve as a natural 
host for graphene quantum dots\cite{GraphaneQdots}
or graphene nanoribbons for nanoroads.\cite{GraphaneGrapheneJunc}
Furthermore, doped graphane has been recently predicted to 
be a high-$T_c$ superconductor.\cite{hTc-SC_graphane}
Such potential application of the graphane structure 
makes it interesting to explore possibilities 
to (locally) modify the the electronic behavior
either by selective hydrogen removal\cite{ErikssonGW,GraphaneQdots,GraphaneGrapheneJunc}
or by geometry-induced band-structure modifications.

The paper is organized as follows.
In Sec.~\ref{sec:Bilayers}, we give a survey of all considered
high-symmetry graphane bilayer configurations.
Section~\ref{sec:Method} presents our computational method.
In Sec.~\ref{sec:Results}, we present, analyze and discuss our results.
Section~\ref{sec:Summary} summarizes our work and contains our conclusions.

\section{High-symmetry graphane bilayers \label{sec:Bilayers}}
\begin{figure}
\includegraphics[width=7.3cm]{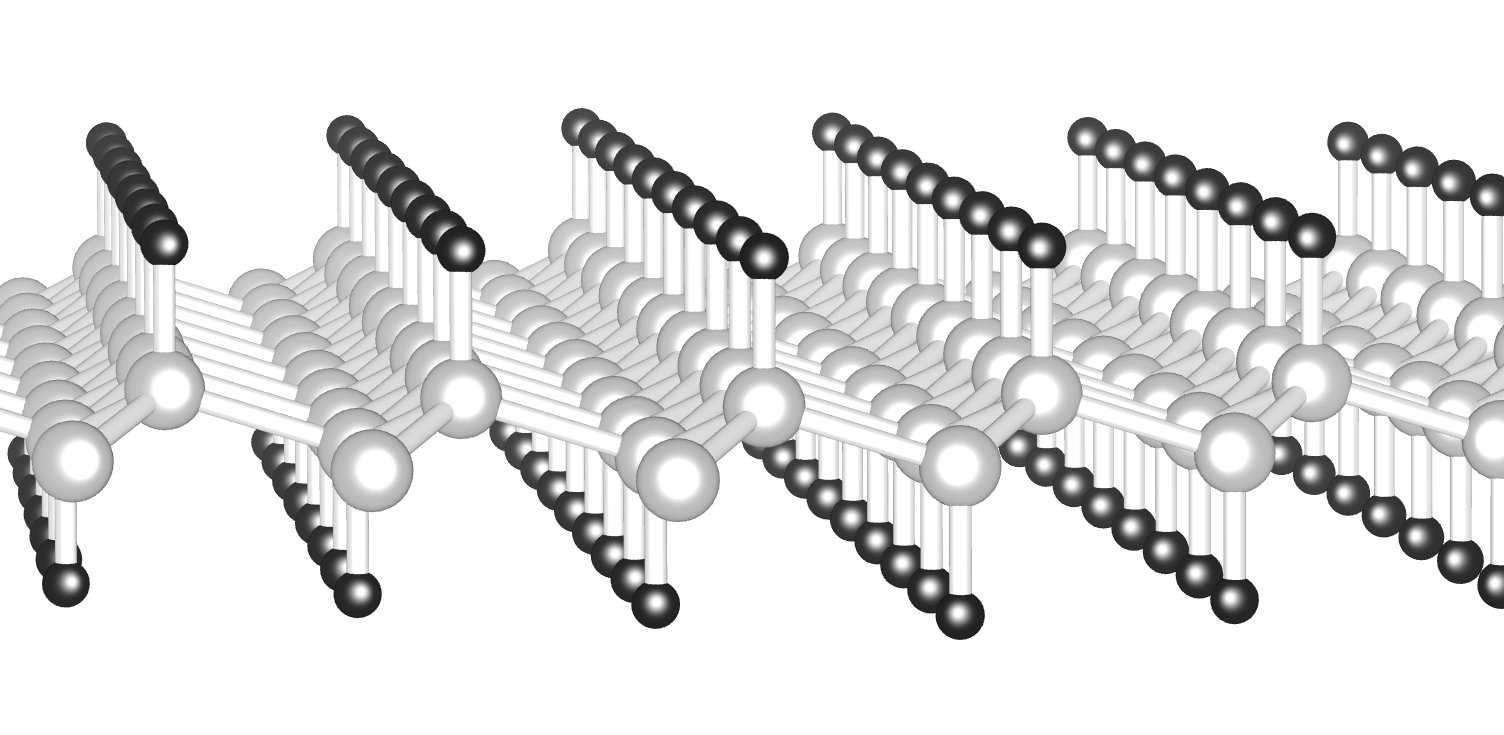}
\begin{tabular}{ccc}
\mc{2}{c}{$\alpha$-type bilayers}&$\beta$-type bilayers\\
\includegraphics[width=2.7cm]{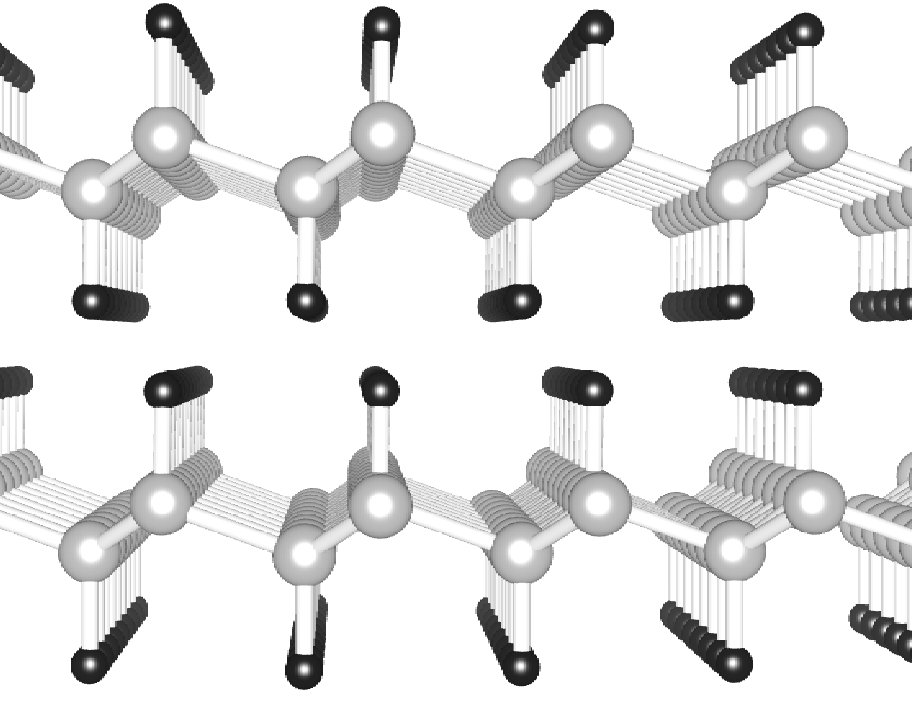}&
\includegraphics[width=2.7cm]{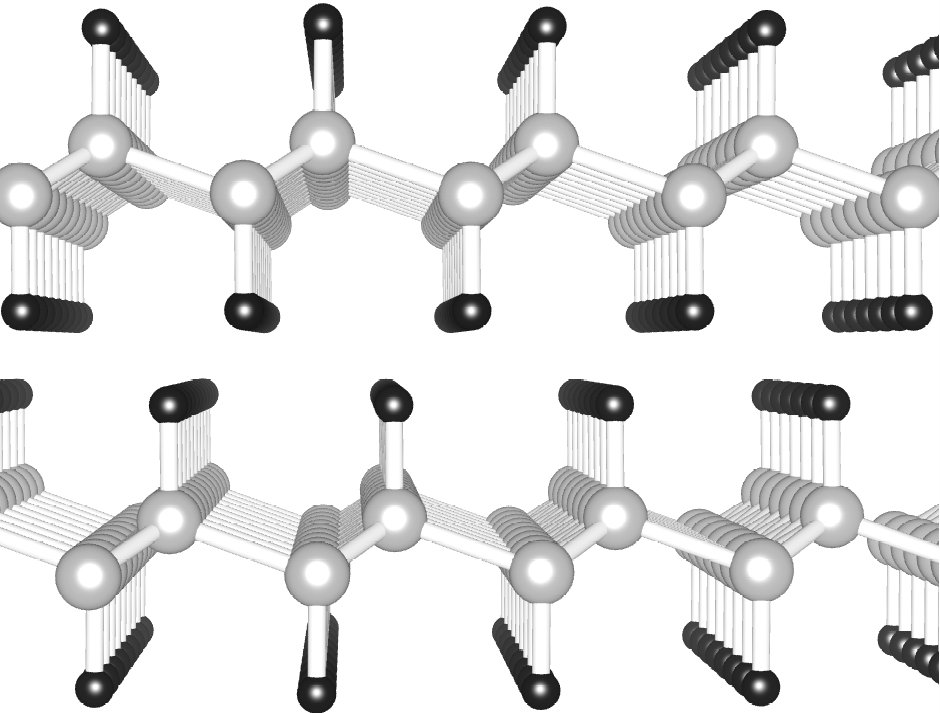}&
\includegraphics[width=2.7cm]{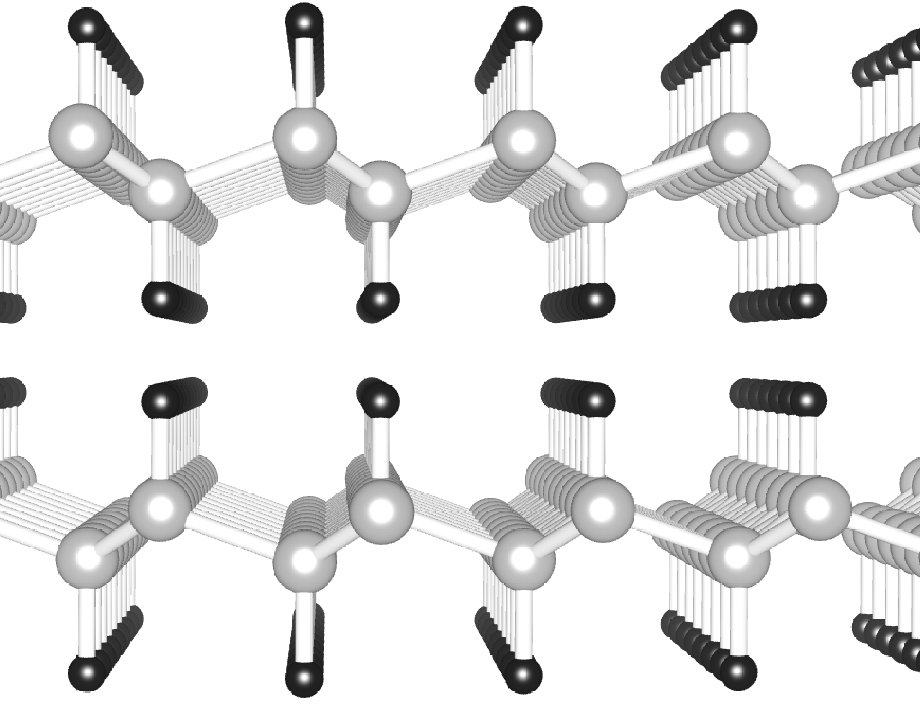}\\
$A_+A_+$&$A_+B_+$&$A_+A_-$\\ [0.2cm]
\includegraphics[width=2.7cm]{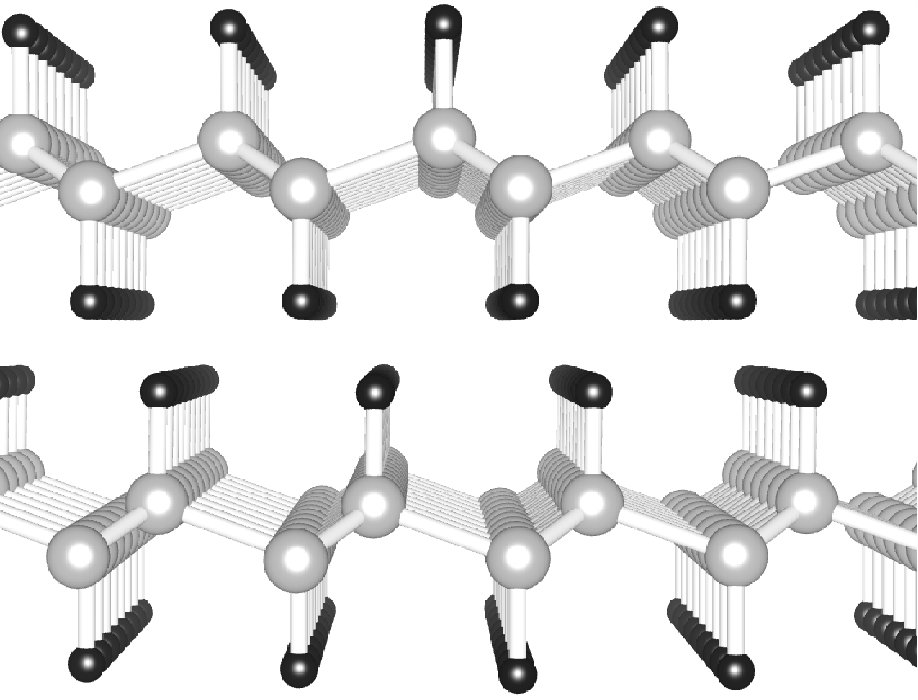}&
\includegraphics[width=2.7cm]{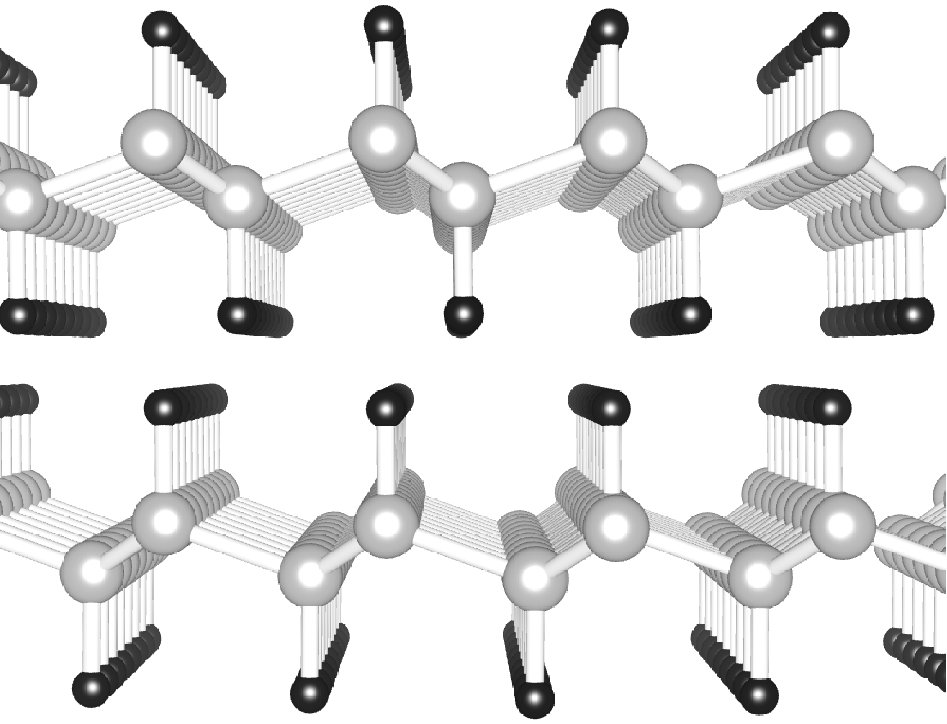}&
\includegraphics[width=2.7cm]{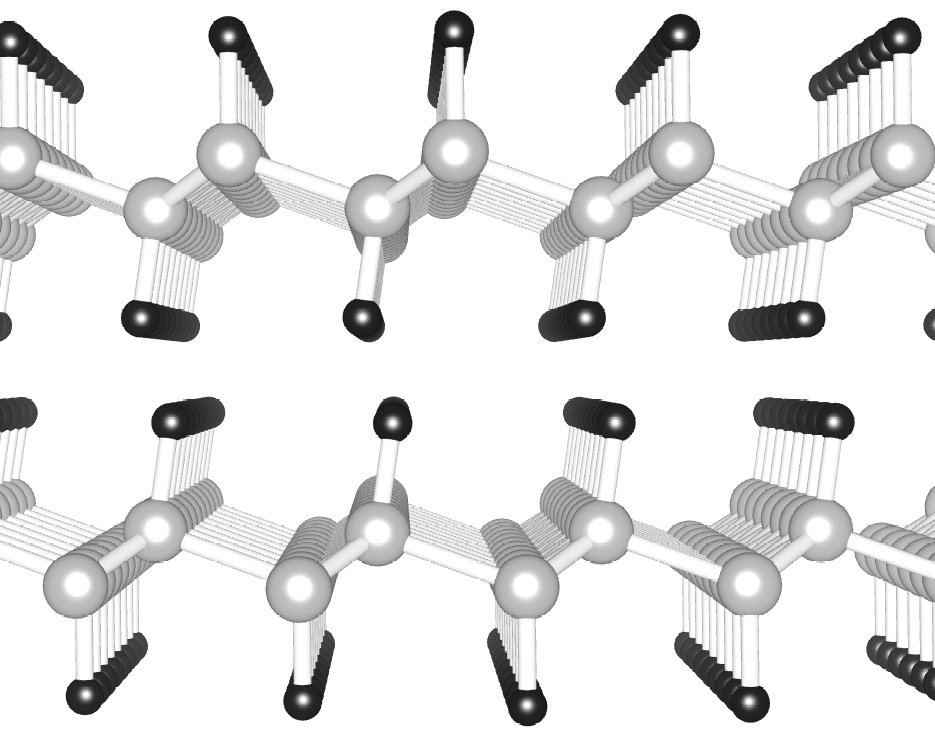}\\
$A_+B_-$&$A_+C_-$&$A_+C_+$
\end{tabular}
\caption{
\label{fig:structures}
Structure of monolayer (top panel) graphane and
of high-symmetry graphane bilayer (BL) configurations.
The labeling  ($A_{+}A_{-}$, \ldots)
provides an unambiguous and exhaustive identification of all
high-symmetry atomic configurations for the BL system
as explained in the text.
Light large spheres represent C atoms,
black small spheres represent H atoms.
}
\end{figure}

The set of lower panels in Fig.~\ref{fig:structures} 
shows all six high-symmetry arrangements of bilayer (BL) graphane.
These can be grouped into two different types. 
In $\alpha$-type BL, 
the graphane sheets are interlocked with each other.
In $\beta$-type BL, 
the H atoms from different graphane sheets
(located between the sheets) 
sit on top of each other.
We calculate and compare all of these configurations 
that make up the $\alpha$- and $\beta$-type sets of stacking configurations.

The set  of different (high-symmetry) arrangements 
for the BL systems are found as follows.
We label the sheets according to the location of vacancy in the C backbone
in the unit cell ($A$, $B$, or $C$ sites).
In addition, the distortion of the C backbone 
along the $z$-direction ($+$ or $-$) of the first occupied C site
(counted along the main diagonal starting from the vacancy stacking)
is indicated as a subscript label.
In all BL, the first layer can be arbitrarily chosen to be an $A_+$ layer.
The second layer is placed on top of the first  one of the following actions:
(i)~copying the bottom layer and moving it along the $z$-direction ($A_+A_+$);
(ii)~flipping the bottom layer and moving it along the $z$-direction ($A_+A_-$);
(iii)~as in (i) and additionally moving it along the long diagonal of the 2D graphane lattice
by one third ($A_+B_+$);
(iv)~as (ii)  and additionally moving it along the  long diagonal of the 2D graphane lattice
by one third ($A_+B_-$);
(v)~as in (i) and additionally moving it along the long diagonal of the 2D graphane lattice
by two thirds ($A_+C_+$);
(vi)~as in (ii) and additionally moving it along the long diagonal of the 2D graphane lattice
by two thirds ($A_+C_-$).

\section{Computational Method \label{sec:Method}}

\subsection{vdW binding}
We map out the energy variation of bilayer (bulk) graphane
as a function of the separation between two graphane sheets (the $c$-parameter of the bulk unit cell)
using (non-selfconsistent) vdW-DF calculations.
For BL, we employ supercells with our optimized 
1$\times$1 graphane in-plane lattice parameters 
[$a_1 =(a,0,0)$, $a_2=(a/2, \sqrt{3}/2\cdot a, 0)$ with $a=2.532$] 
and a height of 30~\AA.
For the bulk, we optimize the $c$-parameter of the periodic unit cell [$a_3=(0,0,c)$]
starting from the optimal value of the BL separation.

Our calculations combine selfconsistent DFT calculations
in the generalized gradient approximation (GGA) 
with three (non-selfconsistent) versions of the vdW-DF method.
The GGA calculations are performed with the 
planewave pseudopential\cite{Pseudopotentials} code \textsf{Dacapo},\cite{Dacapo} 
using  PBE\cite{PBE} for exchange and correlation.
We use a planewave cutoff of $500$~eV
and a  4$\times$4$\times$1 (4$\times$4$\times$2) k-point sampling\cite{KpointSampling}.
The three versions of the vdW-DF method that we use are
(i)~the nonlocal correlation functional of \textit{Dion} \etal\cite{vdWDF} 
in conjunction with revPBE\cite{revPBE} for exchange (vdW-DF1),
(ii)~the same correlation functional but with the exchange part of the C09 functional\cite{C09} 
(vdW-DF1-C09$_{\text{x}}$),
and (iii)~the most recent version of the vdW-DF method, Ref.~\onlinecite{vdWDF2} (vdW-DF2).
The latter version uses the refitted form of the 
PW86 functional (rPW86$_{\text{x}}$)\cite{PW86refitted} for exchange.
We obtain total energies as 
\begin{align}
E^{\text{vdW-DF}}[n]=E_{0}[n]+E^{\text{nl}}_{\text{c}}[n].
\label{eq:vdWDF}
\end{align}
Here, $E^{\text{nl}}_{\text{c}}[n]$ is the energy obtained from one of the non-local functionals 
of Refs.~\onlinecite{vdWDF} and \onlinecite{vdWDF2},
and $E_0[n]$ is given by
\begin{align}
E_{0}=E^{\text{PBE}}_{\text{tot}}-E^{\text{PBE}}_{\text{xc}}+E^{\text{VWN}}_{\text{c}}+
E^{\text{v}}_{\text{x}}.
\label{eq:Enl}
\end{align}
where $E^{\text{VWN}}_{\text{c}}$ is the VWN-LDA\cite{VWN} correlation energy
and the subscript 'v' denotes the version of the exchange functional
(revPBE$_{\text{x}}$, C09$_{\text{x}}$, or rPW86$_{\text{x}}$).
We define the layer binding energy as
\begin{align}
E_{\text{bind}}(d_{\text{cmp}})&=E_{\text{vdW-DF}}(d_{\text{cmp}})-
E_{\text{vdW-DF}}(d_{\text{cmp}}\rightarrow\infty).
\label{eq:Ebind}
\end{align}
Here, $d_{\text{cmp}}$ is the distance between the center-of-mass planes
in each graphane sheet of the monolayer.

Our numerical evaluation of Eq.~(\ref{eq:Ebind})
proceeds in the same way as described in Refs.~\onlinecite{sliding} and
\onlinecite{Method1,Method2,Method3,Method4}.
In particular, because of a small but nonnegligible sensitivity of 
the nonlocal correlation on the exact positioning of atoms with respect to the density grid,
we avoid a direct comparsion of $E^{\text{nl}}_{\text{c}}[n]$
for  BL configurations with different ML separations.
Instead we evaluate the layer-binding energy 
by comparing changes in the nonlocal correlation
arising between the actual configuration and a reference
that keeps the same alignment of atoms and grid points.
Specifically, for all configurations we evaluate
the change in nonlocal correlation as
$\Delta E^{\text{nl}}_{\text{c}}[n]=
E^{\text{nl}}_{\text{c, PQ}}[n]-E^{\text{nl}}_{\text{c, P}}[n]-E^{\text{nl}}_{\text{c, Q}}[n]$.
Here $E^{\text{nl}}_{\text{c, PQ}}[n]$ is the nonlocal correlation energy 
of the full BL configuration (with one ML in P and one in Q)
and  $E^{\text{nl}}_{\text{c, P}}[n]$ ($E^{\text{nl}}_{\text{c, Q}}[n]$)
is the nonlocal energy of the configuration where one 
ML has been removed from Q (P) 
while the other is kept at precisely 
the same location P (Q)
as in the BL configuration.
Further details on our approach to increase the accuracy of vdW-DF 
are provided in Refs.~\onlinecite{sliding} and
\onlinecite{Method1}.

\begin{figure}
\begin{tabular}{c}
\includegraphics[width=8cm]{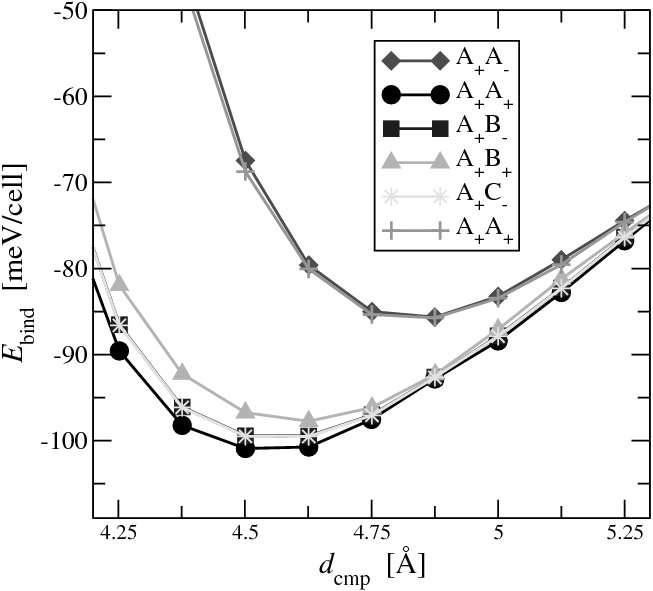}\\
\includegraphics[width=8cm]{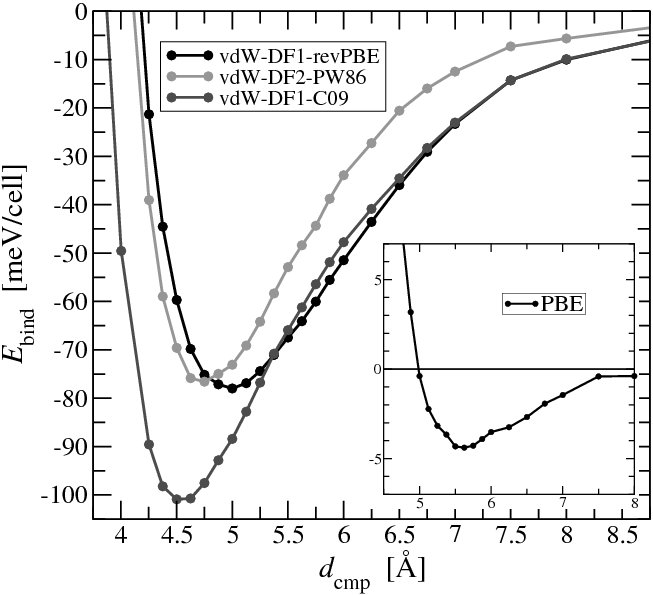}
\end{tabular}
\caption{
\label{fig:vdWDF}
Calculated layer-binding energy variations as functions of 
the center-of-mass (cmp) separation
$d_{\text{cmp}}$ between two graphane sheets.
The top panel compares energy variations of the six different high-symmertry
graphane configurations calculated with  vdW-DF1-C09$_{\text{x}}$.
The $\alpha$-type bilayer have in general lower energy than $\beta$-type configurations.
The A$_+$A$_+$ on-top configuration shows the strongest binding.
The bottom panel compares the different version of vdW-DF
for the A$_+$A$_+$ configuration.
All versions of vdW-DF predict qualitatively the same energy variations
but the detailed numerical values of the binding separations and energies vary.
The insert emphasizes that  GGA calculations provide no meaningful
account of the binding.
}
\end{figure}

\subsection{Band structure}
We determine band structures with pure  GGA calculations
for various k points.
We fix the BL separation (bulk $c$-parameter) to the value calculated with vdW-DF,
so probing indirect effects of vdW binding on electron behavior.
The (selfconsistent GGA) input density for these (non-selfconsistent GGA) band-structure calculations
is obtained using a planewave cutoff of 500~eV and a 20$\times$20$\times$1 (20$\times$20$\times$5)
k-point sampling\cite{KpointSampling} for the bilayer (bulk).

For the bilayer, the Brillouin zone (BZ) is two-dimensional
and relevant k points are $\Gamma=(0,0,0)$, $\text{K}=(2/3, 1/3,0)$
and $\text{M}=(1/2,1/2,0)$.\cite{BZpoints}
All special points are given in units of the reciprocal lattice vectors.
We calculate the band variations along the paths
$\overline{\text{K}\Gamma}$, $\overline{\Gamma\text{M}}$,
and $\overline{\text{K}\text{M}}$.

For the bulk, the BZ is three-dimensional.
Therefore non-zero values of $k_z$ are important
and we also calculate the band variations
along $\overline{\text{H}\text{A}}$, $\overline{\text{A}\text{L}}$,
and $\overline{\text{L}\text{H}}$.
Here, the special points are 
$\text{A}=(0,0,1/2)$, 
$\text{H}=(2/3, 1/3,1/2)$,
and $\text{L}=(1/2,1/2,1/2)$.\cite{BZpoints}

\begin{table}
\begin{ruledtabular}
\begin{tabular}{lrrrr}
\mc{4}{c}{bilayer graphane}\\
 				              &vdW-DF1         &vdW-DF2        &vdW-DF1-C09$_{\text{x}}$ \\
$d_{\text{cmp}}$ [\AA]                        & 5.0 	       & 4.75          &   4.5  \\
$E_{\text{bind}}$ [meV/cell]                  & 78	        & 77 	       &  101   \\
$E_{\text{gap}}$ at $\Gamma$ [eV]             & 3.46  (-0.08)  &  3.53 (-0.01) &   3.61 (+0.07) \\ 
$E_{\text{gap}}$ at K [eV]                    & 11.86 (-0.30)  & 11.75 (-0.41) &  11.61 (-0.55)\\  
$E_{\text{gap}}$ at M [eV]                    & 10.40 (-0.46)  & 10.31 (-0.55) &  10.24 (-0.62)\\[0.2cm] 
\hline
\mc{4}{c}{bulk graphane}\\
 			                       &vdW-DF1         &vdW-DF2        &vdW-DF1-C09$_{\text{x}}$ \\
$c$ [\AA]                                      & 4.8  	        &  4.7          &   4.5     \\
$E_{\text{bind}}$ [meV/cell]                   & 93	        &  94           &   127 \\
$E_{\text{gap}}$  at $\Gamma$ [eV]             &  7.23 (+3.69)  &   7.50 (+3.96)&  8.01 (+4.47)\\
$E_{\text{gap}}$  at K  [eV]                   & 12.77 (+0.61)   &  12.83 (+0.67)& 12.66 (+0.50)\\
$E_{\text{gap}}$  at M [eV]                    & 10.00 (-0.86)   &   9.95 (-0.91)&  9.85 (-1.01)\\
$E_{\text{gap}}$  at A [eV]                    &  3.52 (-0.02)   &   3.62 (+0.08)&  3.88 (+0.34)\\
$E_{\text{gap}}$  at H  [eV]                   & 11.63 (-0.53)   &  11.58 (-0.58)& 11.48 (-0.68)\\
$E_{\text{gap}}$  at L [eV]                    & 11.27 (+0.41)   &  11.27 (+0.41)& 11.28 (+0.42)\\
\end{tabular}
\end{ruledtabular}
\caption{
\label{tab:summary}
Binding separations $d_{\text{cmp}}$ (the $c$- lattice parameter)
layer-binding energies energies $E_{\text{bind}}$, 
and band gaps $E_{\text{gap}}$
at several k points in the Brillouin zone
of bilayer (bulk) graphane.
All quantities are calculated with three versions of vdW-DF.
For the band gaps, the difference with respect 
to a graphane monolayer 
are given in parentheses 
[a negative value corresponds to a 
decreased band gap in the bilayer (bulk)].
}
\end{table}

\begin{figure}
\begin{tabular}{c}
\includegraphics[width=8.5cm]{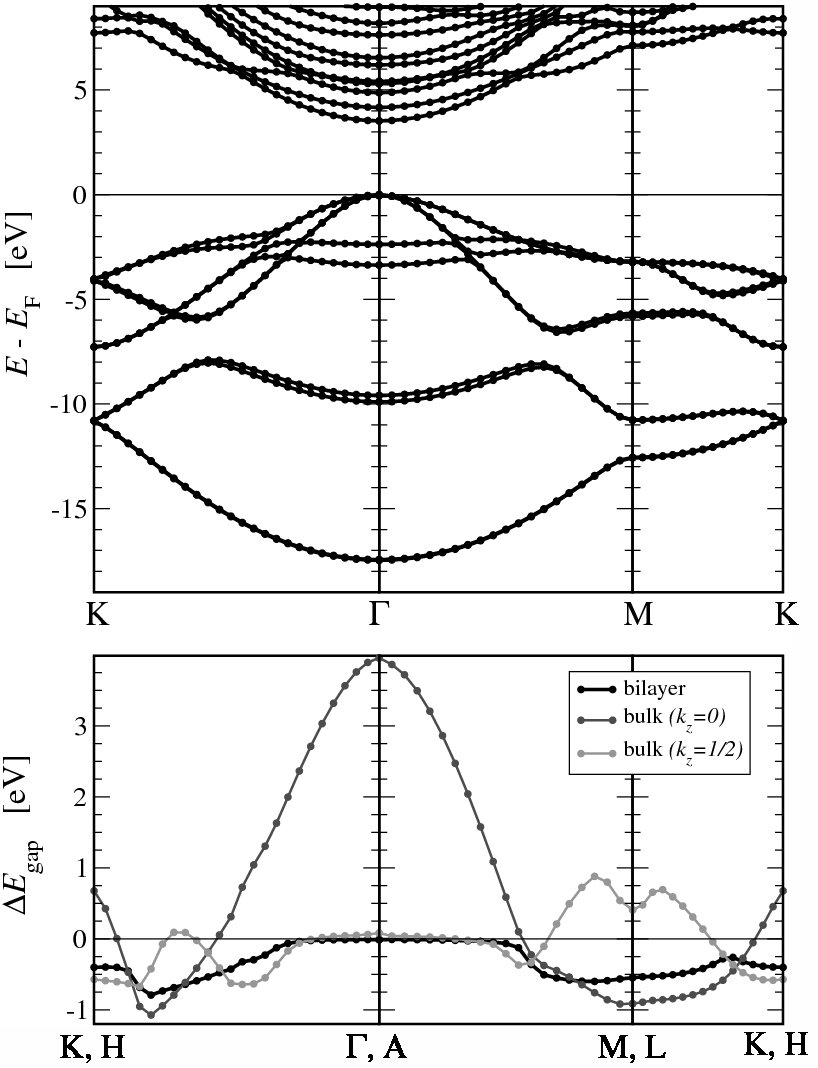}
\end{tabular}
\caption{
\label{fig:BS1}
Electronic band structure of bilayer (BL) and bulk graphane.
The top panel shows the overall band diagram 
along  $\overline{\text{K}\Gamma}$, $\overline{\Gamma\text{M}}$, and $\overline{\text{M}\text{K}}$
of a $A_+A_+$ BL at the binding  separation predicted by vdW-DF2.
Apart from the fact that each  band exists as a pair of bands, 
the BL band-structure qualitatively agrees with that of 
of a monolayer (ML, see Fig.~3 in Ref.~\onlinecite{ref:Sofo_Graphane_2007}).
The bottom panel shows the $k$-dependent (direct) band-gap variation
in the BL and bulk with respect to the band-gap variation in the ML.
For the bulk, we also show the band-gap variation
along $\overline{\text{H}\text{A}}$, $\overline{\text{A}\text{L}}$,
and $\overline{\text{L}\text{H}}$.
In the BL, at and around  the $\Gamma$ point,
the gap is essentially unchanged.
However, away from the $\Gamma$ point
significant band-gap reduction is observed.
In the bulk, local band-gap modifications can be both positive
and negative and their absolute can be even larger
than in the BL.
}
\end{figure}

\section{Results: Predicted properties of bilayer and bulk graphane \label{sec:Results}}

Figure~\ref{fig:vdWDF} shows the calculated variations in layer-binding energies
as  functions of the separation between the center-of-mass planes ('cmp') 
of the two graphane monolayers (ML) in a bilayer (BL).
The top panel compares the energy variations for
the configurations with different stackings
using  vdW-DF1-C09$_{\text{x}}$.
The energy variations split according to the grouping into 
$\alpha$- and $\beta$-type configurations.
The $\alpha$-type configurations have a smaller binding separation
and a higher binding energy;
the $A_+A_+$ stacking shows the strongest bonding.\cite{PES}

The bottom panel compares the energy variations of the $A_+A_+$ BL
for vdW-DF1, vdW-DF1-C09$_{\text{x}}$, and vdW-DF2.
Qualitatively, all functionals yield the same energy variations.
The insert shows the energy variation for the
$A_+A_+$ configuration obtained from pure PBE calculations
and illustrates that no meaningful binding is predicted 
without an account of vdW forces.

Table~\ref{tab:summary} lists and compares numerical results 
for the calculated binding separations 
and layer-binding energies for the BL.
The binding separations and energies 
range from  4.5~\AA\ to 5.0~\AA\ 
and 75~meV/cell to and 102~meV/cell,
depending on the version of vdW-DF.\cite{LDAcomment}
The binding energy is comparable to that in a graphene BL\cite{ref:GraphitevdW}
(94~meV using vdW-DF1).

Table~\ref{tab:summary} also lists the calculated  lattice constant $c$ 
and the corresponding layer-binding energies for 
a fictitious bulk crystal of graphane.
It is possible that such a 3D graphane system might eventually be synthesized.
We here present predictions of the expected structure,
using our analysis of stacking
in the BL as starting point.
In particular we focus on $A_+A_+$ stacking and find 
that the lattice constant essentially coincides with $d_{\text{cmp}}$ in the bilayer.
The binding energy is slightly increased in the 
bulk and varies between 93~eV and 127~meV.
These numbers also compare to the binding in graphite\cite{ref:GraphitevdW}
(100~eV using vdW-DF1).

The top panel of Fig.~\ref{fig:BS1} presents the overall PBE band-diagram
for the $A_+A_+$ stacked graphane BL at the binding separation predicted by vdW-DF2.
Corresponding band diagrams at vdW-DF1 or vdW-DF1-CO$_{\text{x}}$ binding separations 
are qualitatively similar.
Apart from the fact that each band occurs as a pair of bands,
the band structure also agrees qualitatively
with that of the ML  (see, for example, Ref.~\onlinecite{ref:Sofo_Graphane_2007}).

The bottom panel of Fig.~\ref{fig:BS1} summarizes some differences
between the BL (bulk) and the ML,
documenting the changes occurring in the $k$-dependent
band gap with the BL (bulk) formation.
We plot the differences $\Delta E_{\text{gap}}^{\text{BL/bulk}}(k)=E_{\text{gap}}^{\text{BL/bulk}}(k)-E_{\text{gap}}^{\text{ML}}(k)$
along $\overline{\text{K}\Gamma}$, $\overline{\Gamma\text{M}}$, and $\overline{\text{M}\text{K}}$.
In addition, for the bulk, we also plot
$\Delta E_{\text{gap'}}^{\text{bulk}}(k)=E_{\text{gap}}^{\text{bulk}}(k')-E_{\text{gap}}^{\text{ML}}(k)$.
Here, $k'$ is along 
$\overline{\text{H}\text{A}}$, $\overline{\text{A}\text{L}}$,
and $\overline{\text{L}\text{H}}$ in the bulk;
$k$ is the corresponding k point along
$\overline{\text{K}\Gamma}$, $\overline{\Gamma\text{M}}$, and $\overline{\text{M}\text{K}}$
in the ML (and therefore $k_x=k'_x$, $k_y=k'_y$, while $k_z=0$ and $k'_z=1/2$).
A summary of the numerical values of band gaps at 
the special points  (calculated with various choices of vdW-DF and corresponding BL binding separations 
or bulk lattice constants $c$)
and their deviations from the corresponding values in the ML is given
in Table~\ref{tab:summary}.

We find large modifications of the band structure,
\textit{indirectly} induced by the vdW interactions and
summarized by the $k$-dependent band-gap differences.
In the BL, the direct band gap can deviate by up to $\sim 0.8$~eV (between K and $\Gamma$) 
with respect to the ML gap (see bottom panel of Fig.~\ref{fig:BS1}).
In the bulk, deviations can be as large $\sim-1.2$~eV ($\sim+4$~eV)
in some regions of the Brillouin zone (BZ)
near the H point ($\Gamma$ point).

\section{Discussions: Bilayer graphane}
Focusing on the direct band gap in BL graphane,
we find modifications that are strongly $k$-dependent.
At the K and M points (and in other regions),
the modifications are significant.
At the $\Gamma$ point, where the gap is smallest in the ML (and in the BL),
no modifications occur,
rendering the BL system electronically similar to the ML system.
Nevertheless, qualitative understanding of the origin of
the different modifications in the various regions
is important to gain further insight into the relevance
of vdW interactions for materials band-structure.

In the following, we explore the band-gap modifications 
upon formation of graphane BLs  in more detail.
We focus on the \textit{indirect, geometry-induced
effects} of vdW binding outlined in Fig.~\ref{fig:ChargeRearrangements}:
the effect of a modified electronic environment in the region between
the MLs that form the BL arising from a superposition of
two ML electron densities (see left panel);
the effect of self-consistent charge rearrangements (on the GGA level)
of this superpositioned density (see right panel);
the effect of potential hybridization of wave functions (WFs) (not shown in the figure).

Our analysis suggests that the band-gap modifications
should be interpreted as a concerted interplay between
WF hybridization and electrostatic interaction between 
the hybridized WFs with the modified environment.
The relevant WFs are unoccupied conduction-band (CB) WFs.
Self-consistent (SC) charge rearrangements 
(with respect to the superposition of ML densities)
do not play a significant role.

\begin{figure}
\includegraphics[width=8cm]{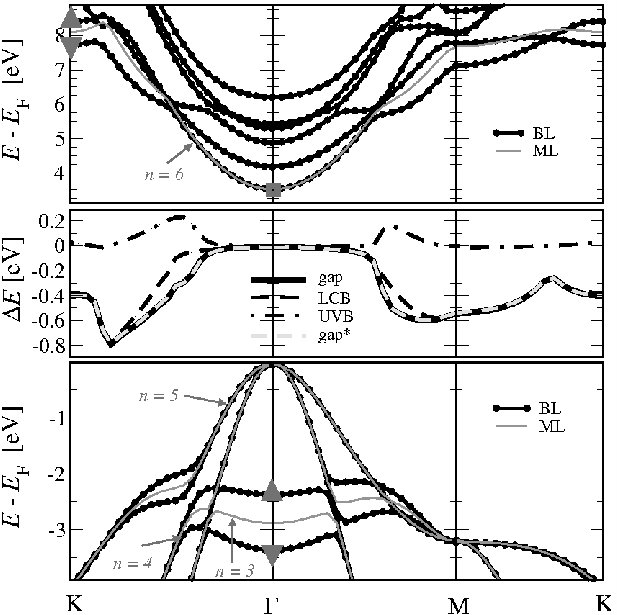}
\caption{
\label{fig:SC}
Band-origin of gap modifications, self-consistent charge rearrangements and wave function hybridization.
The top and bottom panels show the band structure in the LCB and UVB respectively.
ML graphane bands have been included to highlight that,
in the BL, bonding and antibonding hybrid bands are formed.
For the MLs, the band index $n$ is annotated.
Also, we have marked some combinations of bands and k points for which wavefunction 
will be analyzed later.
The mid-panel shows the variation of direct band-gap differences between the BL and ML (solid black),
the variation of UVB energy differences (dashed-dotted black),
and the variation of LCB energy differences (dashed black).
We also show the variation of band-gap differences calculated 
from the non-selfconsistent BL charge density 
obtained by adding two ML densities (dashed light gray).
}
\end{figure}

\subsection{Band-origin of gap reduction}
In the mid-panel of Fig.~\ref{fig:SC}, we show that the main contribution to the 
observed band-gap variation comes from the lower conduction-band (LCB).
We plot the variations of the upper valence-band (UVB) energies (dashed-dotted curve)
and of the LCB energies (dashed curve),
\begin{align}
\Delta E_{\text{XXX}}(k)=\left(E_{\text{XXX}}^{\text{BL}}(k)-E_{\text{F}}^{\text{BL}}\right)-
\left(E_{\text{XXX}}^{\text{ML}}(k)-E_{\text{F}}^{\text{ML}}\right).
\end{align}
Here $E_{\text{XXX}}^{\text{BL/ML}}(k)$ is the energy of the highest (occupied)
valence band (XXX = UVB)
or the energy of the lowest (unoccupied)  conduction band (XXX=LCB)
at $k$ in the BL/ML
and $E_{\text{F}}^{\text{BL/ML}}$ is the Fermi level
in the BL/ML.
(The Fermi level is here defined by the energy of the highest occupied state).

We find that $|\Delta E_{\text{LCB}}(k)|$ is typically much larger than 
$|\Delta E_{\text{UVB}}(k)|$.
The latter is essentially zero.
In the UVB, nonzero contributions to the band-gap modifications
are only found in small regions.
As can be seen from the bottom panel of Fig.~\ref{fig:SC}
where we contrast BL VBs with ML VBs,
the topmost BL UVBs are non-degenerate there, 
indicating  WF hybridization.
Nevertheless, in regions where $|\Delta E_{\text{gap}}(k)|$ is large,
only $|\Delta E_{\text{LCB}}(k)|$ contributes.
Thus, we assign the modifications of the band-gap variation in the BL
primarily to the modifications in the LCB energy variation.

\subsection{Self-consistent charge rearrangements}
In the mid-panel of Fig.~\ref{fig:SC} we also contrast 
band-structure modifications
obtained from the SC BL charge density
with band-structure modifications obtained 
from the non-SC charge density, 
constructed as a superposition of ML densities.
The solid black line (gap) corresponds to the SC case,
the dashed light-gray line (gap$^*$) corresponds to the non-SC case.
At the displayed resolution, the curves cannot be distinguished.
We find that the differences between both band-gap variations are at the meV level.
Thus, the charge rearrangements\cite{ChargeTransfer}
shown in the right panel of Fig.~\ref{fig:ChargeRearrangements}
do not appreciably contribute to the band-gap variation.
In fact, this also applies for the overall band structure variation.

Further charge rearrangements\cite{Thonhauser2007} that are inherent to vdW forces
(and require a SC vdW-DF calculations)\cite{Thonhauser2007,Soler, GPAW}
are expected to be even smaller than those resulting at the the GGA level.
The inherent  rearrangements are not expected to be of importance for the band structure.
This  justifies our use of  non-SC vdW-DF in this study of bilayer graphane.

\subsection{Hybridization and concerted effects on kinetic and potential energy}
In the remainder of this section we investigate the 
role of hybridization for the observed band-structure modifications.
In the simplest picture of hybridization
a bonding and antibonding  hybrid WF
can be formed when two degenerate atomic or layer WFs  $\phi_1$ and $\phi_2$ 
approach each other,
\begin{align}
\psi^{\pm} = \frac{1}{\sqrt{2}}(\phi_1\pm\phi_2).
\label{eq:SimpleHybrid}
\end{align}
If the actual BL WF equals $\psi^{+}$
or alternatively $|\psi_{\text{BL}}|^2=|\psi^{+}|^2$,
its energy is shifted to lower energies.
Similarily, if the actual BL WF equals 
$\psi^{-}$
or alternatively $|\psi_{\text{BL}}|^2=|\psi^{-}|^2$,
its energy is shifted to lower energies.
The energy shifts 
in such a simple picture of hybridization effects
are due to a gain and a loss of \textit{kinetic} energy.

This simple hydrogen-like picture of hybridization needs to be modified
in the present case for two reasons.
First, the unhybridized ML WFs are already complex objects
possessing internal nodes.
Second, the WFs live in a background effective \textit{potential} $V$.
The hybrid WFs will then interact with this potential
leading to further modifications of 
the actual bonding (B) and actual antibonding (A) WFs
$\psi^{\text{B}}$ and $\psi^{\text{A}}$.
The actual hybrid WFs  $\psi^{\text{B}}$ and $\psi^{\text{A}}$
will therefore no longer coincide with $\psi^{+}$ and 
$\psi^{-}$,
nor will their energy shift be only of kinetic nature.

We now move the discussion to a comparison of ML and BL graphane.
For the ML,  we denote the WFs by $\phi_{n,\textbf{k}}$,
where $n$ is the band index and $\textbf{k}$ the wave vector.
For the BL, $\psi_{n,\textbf{k}}^{\text{B}}=\psi_{2n-1,\textbf{k}}$
is the bonding WF associated with two $\phi_{n,\textbf{k}}$
located on different sheets;
$\psi_{n,\textbf{k}}^{\text{A}}=\psi_{2n,\textbf{k}}$
is the antibonding WF.

These WFs (here collectively denoted by  $\varphi_{n,\textbf{k}}$)
satisfy the Kohn-Sham equation 
\begin{align}
\left[-\nabla^2+(V_{\text{eff}}-E_{\text{F}})\right]\varphi_{n,\textbf{k}}
=(E_{m, \textbf{k}}-E_{\text{F}})\varphi_{n,\textbf{k}}.
\end{align}
Here, $V_{\text{eff}}$ is the effective potential (which, in general, is different for the ML and BL system),
$E_{\text{F}}$ is the Fermi level (which may also be different in the ML or BL system)
and $E_{n, \textbf{k}}$ the band energy of the WFs (also different in general).
Accordingly, we can separate the kinetic- and potential-energy shifts
of hybrid WFs as
\begin{align}
\Delta T_{n,\textbf{k}}^{\text{B/A}}&=
\langle\psi_{n,\textbf{k}}^{\text{B/A}}|-\nabla^2|\psi_{n,\textbf{k}}^{\text{B/A}}\rangle
- \langle \phi_{n,\textbf{k}}|-\nabla^2|\phi_{n,\textbf{k}}\rangle
\label{DeltaT}\\
\Delta V_{n,\textbf{k}}^{\text{B/A}}
&=\langle\psi_{n,\textbf{k}}^{\text{B/A}}|V_{\text{eff}}^{\text{BL}}-E_{\text{F}}^{\text{BL}}|\psi_{n,\textbf{k}}^{\text{B/A}}\rangle\nn\\
&- \langle \phi_{n,\textbf{k}}|V_{\text{eff}}^{\text{ML}}-E_{\text{F}}^{\text{ML}}|\phi_{n,\textbf{k}}\rangle.
\label{DeltaV}
\end{align}

In the present analysis we focus on a quantitative evaluation of
the per-orbital potential-energy shifts in Eq.~(\ref{DeltaV})
and on a qualitative account of the changes in the kinetic-energy term in Eq.~(\ref{DeltaT}).
We replace the effective potential by the electrostatic potential $V_{\text{es}}$
(consisting of the Hartree potential and the atomic core potentials),
neglecting effects from exchange and correlation,
and give  qualitative accounts of the changes in kinetic energies.
A quantitative comparison of kinetic-energy shifts would be desirable but,
since we are using pseudo-WFs,
the evaluation of Eq.~(\ref{DeltaT}) is
nontrivial\cite{Pseudopotentials}
and beyond the present scope.

We obtain a qualitative analysis of
the changes in kinetic energies 
by plotting the change of partial electron density associated
with an (anti-) bonding BL WF
\begin{align}
\rho_{n,\textbf{k}}^{B/A}=|\psi_{n,\textbf{k}}^{B/A}|^2
\label{eq:rhoAB}
\end{align}
with respect to a sum of or difference between the
corresponding ML WFs,
\begin{align}
\rho_{n,\textbf{k}}^{\pm}=|\psi_{n,\textbf{k}}^{\pm}|^2=1/2|\phi_{n,\textbf{k}}^{\text{ML}_1}\pm\phi_{n,\textbf{k}}^{\text{ML}_2}|^2.
\label{eq:rhoPM}
\end{align} 
The differences
\begin{align}
\rho_{n,\textbf{k}}^{B/A}-\rho_{n,\textbf{k}}^{\pm}
\end{align}
measure the extent to which 
the BL WFs experience
a reduction or enhancement of kinetic energy
with respect to a simple hybridization,
Eq.~(\ref{eq:SimpleHybrid}).

\begin{table}
\begin{ruledtabular}
\begin{tabular}{rrrrrr}
k        & n     &  $\Delta E^{\text{B}}$& $\Delta E^{\text{A}}$ &  $\Delta V^{\text{B}}$ & $\Delta V^{\text{A}}$ \\
         &       &  \multicolumn{4}{c}{in eV}\\
$\Gamma$ & 2 (VB)& -0.17 & 0.14 & 0.20 & -0.34 \\
$\Gamma$ & 3 (VB)& -0.48 & 0.51 & 0.43 & -1.04 \\
K        & 6 (CB)& -0.37 & 0.31 & 0.37 & -0.51 \\
$\Gamma$ & 6 (CB)&  0.00 & ---  &-1.07&  ---
\end{tabular}
\end{ruledtabular}
\caption{
\label{tab:EnergyShifts}
Total energies shifts $\Delta E$ and electrostatic contributions $\Delta V$ 
according to Eqs.~(\ref{DeltaV})
of hybrid WFs in the BL with respect to ML WFs
at $\Gamma$ and K for several bands ($n$ specifies the corresponding ML band
and CB or VB whether this band belongs to the conduction or valence band).
}
\end{table}

\subsection{Hybridization in the valence band}
Hybridization in the valence band is found,
for example, around the $\Gamma$ point in the bands
that correspond to band no 2 and 3 in the ML,
see Figs.~\ref{fig:BS1} and~\ref{fig:SC}.
We emphasize that the hybridization of the corresponding WFs 
is not a signature of binding.
The energy splits are (essentially) symmetric and 
since both the bonding and the antibonding states are occupied,
there is no (significant) net gain in total energy.

In Table~\ref{tab:EnergyShifts},
we list the total-energy shifts $\Delta E^{\text{B/A}}$ (obtained directly from our calculations)
and the potential-energy contributions to these shifts
$\Delta V^{\text{B/A}}$ for the corresponding WFs.
Interestingly, we find that the electrostatic contributions
to the energy shifts are positive for bonding BL WF 
whereas they are negative for the antibonding BL WF.
The kinetic-energy gain (loss) must therefore be significantly larger than the loss (gain)
in potential energy of the bonding (antibonding) BL WF
to produce the ordering shown in the bottom panel of Fig.~\ref{fig:SC}
(compare states identified by triangles at the $\Gamma$ point).

\begin{figure}[t]
\begin{tabular}{c}
\includegraphics[width=8.4cm]{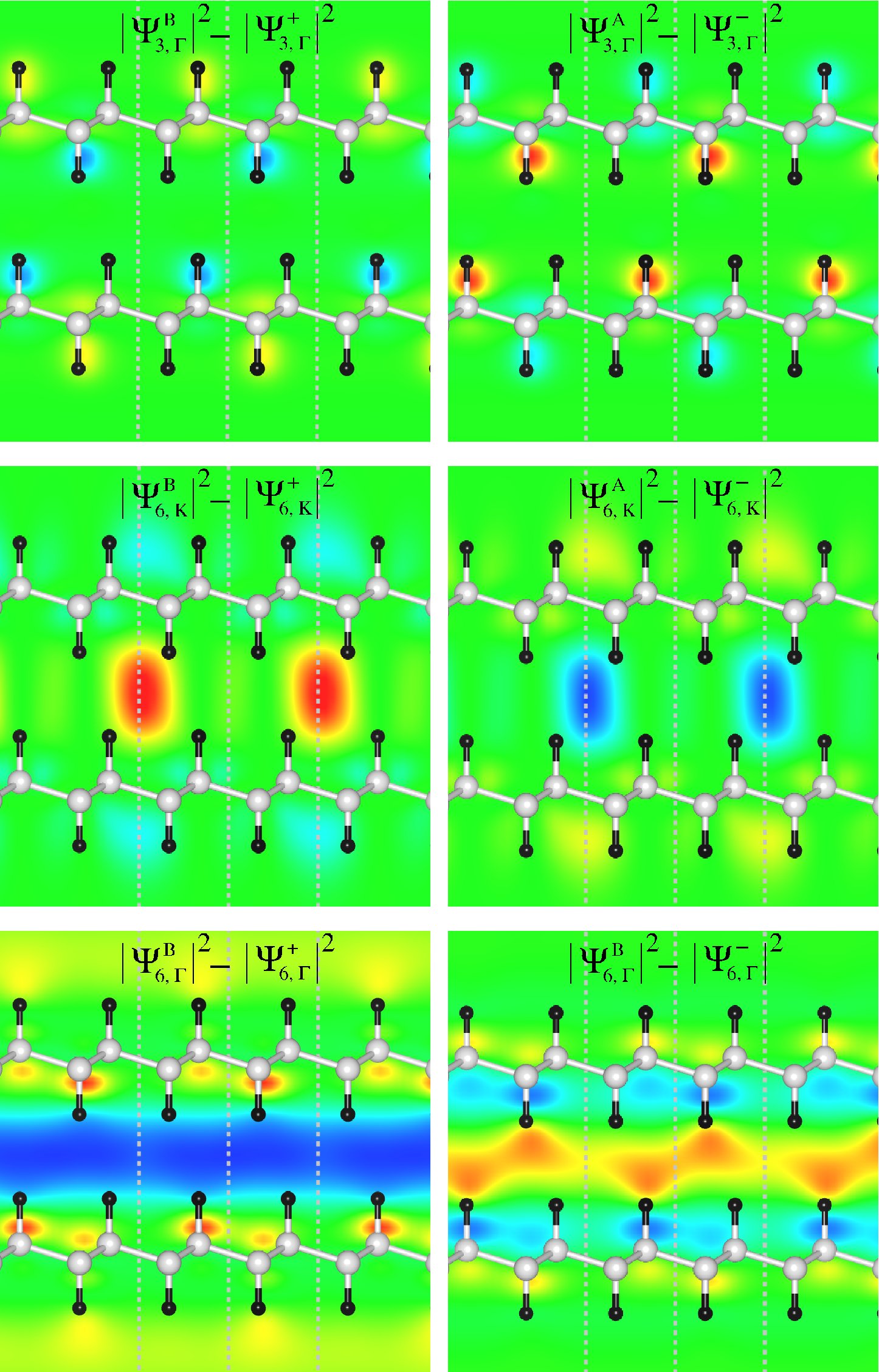}
\end{tabular}
\caption{
\label{fig:WFs}
(Color)
Qualitative analysis of kinetic-energy contributions to total-energy shifts
of hybrid WFs in terms of contour-plot sections of differences
between partial electron densities of bonding (antibonding) BL WF 
and sums of (differences between) associated ML WFs, Eqs.~(\ref{eq:rhoAB}) and (\ref{eq:rhoPM}).
Positive (negative) contours are in red (blue).
A green color between the MLs in the top panel reflects 
that the BL VB WFs at $\Gamma$
are essentially simple hybridizations of the ML WFs.
In the mid-panel, our comparison indicates that the 
bonding (antibonding) LCB BL WFs at K possess
a lower (higher) kinetic energy than the
simple bonding-type sum of (antibonding-type difference between)
ML WFs.
The bottom panels show that the lowest CB BL WF at $\Gamma$
differs significantly from both the 
sum of and the difference between ML WFs
and that the BL WF possesses a more complex hybridization nature 
than implied by a simple picture.
However, the closer resemblance to the antibonding difference
of ML WFs indicates a loss of kinetic energy with respect to an 
individual ML WF.
}
\end{figure}

The top panels of Fig.~\ref{fig:WFs} show
sections (through the main diagonal of the unit cell)
of the difference between the partial densities that correspond to the  bonding 
and antibonding  VB BL WFs at $\Gamma$
and the partial densities that correspond to the pure sum of 
(difference between) the corresponding ML WFs (with band index 3).
The green color (indicating an absence of any decrease or enhancement
relative to a simple hybridization)
between the MLs in both panels shows that 
the BL WFs gains or looses kinetic energy.
These kinetic-energy changes evidently more than make 
up the concerted changes in the
potential-energy terms.

\subsection{Conduction-band modifications}
For the CB,
we focus on the band-gap modifications 
and WFs at the points marked in the top panel of Fig.~\ref{fig:SC}
(at K and $\Gamma$).
At K, the total-energy splitting leads to the reduced band gap.
As shown in Tab.~\ref{tab:EnergyShifts},
the potential-energy shift is positive
for the bonding BL WF at K (marked with a downward triangle in Fig.~\ref{fig:SC}) 
and negative for the antibonding BL WF (marked with an upward triangle).
This observation is in line with those made at $\Gamma$
in the VB.

The mid-panels of Fig.~\ref{fig:WFs} show 
the differences between partial density associated with the  lowest bonding (next-lowest antibonding)
CB BL WFs and the density associated with the pure sum of 
(difference between) the corresponding ML WFs at K.
For the bonding BL WF, the partial density is increased
between the two ML with respect to the pure sum of ML WFs.
For the antibonding BL WF, the partial density is 
decreased with respect to the difference between
the ML WFs.
This indicates that the kinetic energy gain (loss)
of the bonding (antibonding) BL WF is larger
than within a simple hybridization picture
where the hybrid WFs already possessing a kinetic energy
gain (loss) with respect to an individual ML WF.

At $\Gamma$, the contribution of the potential energy to the total shift
of the lowest lying CB WF (marked with a square Fig.~\ref{fig:SC})
is negative, see Table~\ref{tab:EnergyShifts}.
In a simple hybridization picture
one would expect a WF with a bonding nature
and an additional decrease of kinetic energy.
The vanishing shift in total energy,
however, requires a kinetic-energy offset
that compensates for the negative potential-energy shift.
Specifically, one must therefore expect 
a more complicated hybridization of this BL WF.

The bottom panels of Fig.~\ref{fig:WFs} show
the difference between the partial density of the lowest CB BL WF
and the pure sum of (left panel) and the pure difference between (right panel)
the corresponding ML WFs at $\Gamma$.
The significantly negative value of the contours 
between the MLs in the left panel 
indicates that the BL WF at $\Gamma$
has a higher kinetic energy than the 
pure sum of ML WFs.
Similarly, the kinetic energy is lower than
in the difference between ML WFs,
indicated by the positive value
of the contours between the MLs in the right panel.

The contour plots show that the lowest CB BL WF at $\Gamma$ 
is not a simple hybridization of ML WFs.
Also, we notice that the differences in the right panel
are not as pronounced as the differences in the left panel.
This suggests that the lowest CB BL WF at $\Gamma$ 
possesses rather an antibonding nature
(although the lowest CB BL WF has a smaller kinetic energy
than a simple antibonding hybridization).
Such an antibonding nature is consistent with the
vanishing total-energy shift and the negative
potential-energy shift.

\begin{figure*}
\begin{tabular}{c}
\includegraphics[width=17cm]{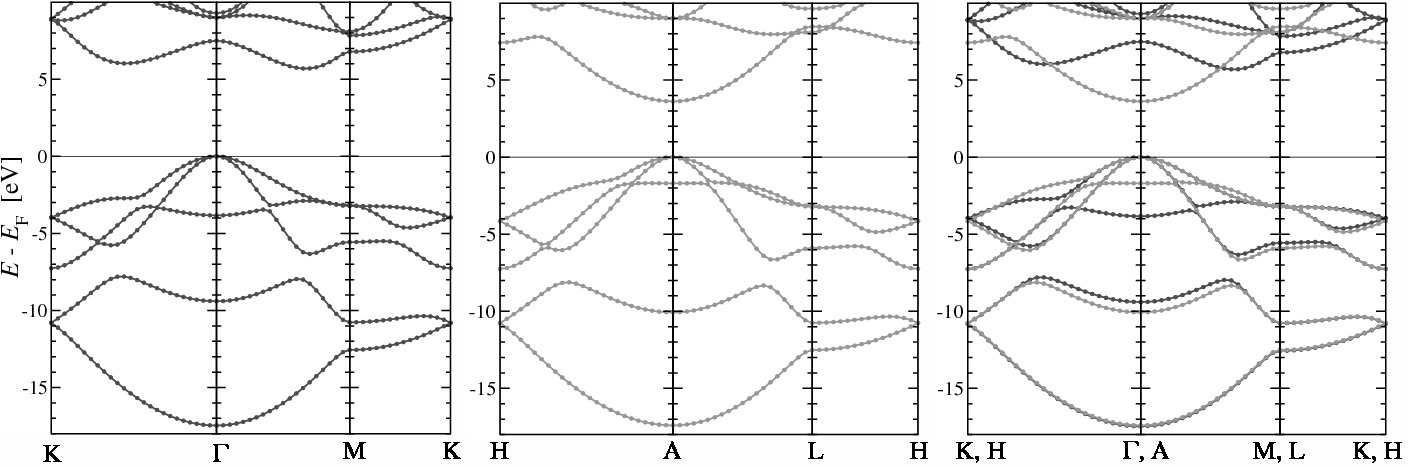}
\end{tabular}
\caption{
\label{fig:BS3}
Band diagrams for bulk graphane.
The left panel shows the band diagram along the same path in k space
as shown for the bilayer in the top panel of Fig.~\ref{fig:BS1}.
The midpanel shows the band diagram along a parallel path in k space
with $k_z=1/2$.
In the right panel we combine both band diagrams 
into one plot and reproduce a band diagram that possess essentially all 
features that are present in the BL band diagram,
compare top panel of  Fig.~\ref{fig:BS1}.
}
\end{figure*}

\section{Discussions: Bulk graphane}
The most pronounced electronic difference between bulk graphane  
and ML or BL graphane is the fact that the A point takes over the 
role of the $\Gamma$ point, see Fig.~\ref{fig:BS1}.
At A, the band gap is smallest in the bulk and the size of that 
gap coincides essentially with the size of the gap at $\Gamma$ in the ML or BL.
Also, in the bulk, the deviations in the band-gap variations 
can be considerably larger than those in the BL, see bottom panel of Fig.~\ref{fig:BS1}.
Here, we give a more detailed analysis of the overall band structure
effects associated with graphane assembly into bulk.

In the the left panel of Fig.~\ref{fig:BS3},
we show the band diagram along 
$\overline{\text{K}\Gamma}$, $\overline{\Gamma\text{M}}$, and $\overline{\text{M}\text{K}}$.
Focusing on the CB, the bulk band structure is very different 
from that of the BL along the same path, see top panel of Fig.~\ref{fig:BS1}.
In particular at the $\Gamma$ point, that gap is approximately twice as large
as in the BL.
Also, the band energy at K is considerably higher
than in the BL.
Only at M, we find similar band energies.
Thus, the effect of vdW-bonding on local features of the band structure
and related observable properties can be dramatic.

In the mid-panel of  Fig.~\ref{fig:BS3},
we show show the band diagram along 
$\overline{\text{H}\text{A}}$, $\overline{\text{A}\text{L}}$,
and $\overline{\text{L}\text{H}}$.
Again, the the bulk band structure is very different 
from that of the BL along the parallel path 
along $\overline{\text{K}\Gamma}$, $\overline{\Gamma\text{M}}$, and $\overline{\text{M}\text{K}}$.
The band energy at H is lower than at L,
while it is the other way around at K and at M in the BL.
Only around the $A$ point 
(corresponding to $\Gamma$ in the BL and ML cases)
do the bulk and ML band diagram 
show similar features.

In the right panel of Fig.~\ref{fig:BS3},
we effectively combine the two diagrams from the left and mid-panel,
calculating the band diagram for a would-be 2-layer unit cell
at $k_z=0$.
The resulting band diagram shows the features of 
the BL band diagram
and of the VB there is also a very good quantitative agreement.
Therefore, the BL band-diagram can partly be understood 
as a zone-folded version of the bulk band-diagram.

\section{Summary and Conclusion \label{sec:Summary}}
This paper predicts and characterizes band-structure modifications
produced indirectly as a geometry-induced effect of
pure dispersive binding between two macromolecules
in  possible new material systems: bilayer and bulk graphane.\cite{BLGraphane1,BLGraphane2}
Using non-selfconsistent vdW-DF calculations,\cite{vdWDFreview, Thonhauser2007, vdWDF, vdWDF2, C09}
we determine the binding separation ($c$-lattice parameter) 
and binding energies in vdW-bonded graphane bilayers (bulk).
We use the calculated separations to obtain corresponding GGA-DFT band diagrams.
Our results demonstrate that vdW interactions can significantly 
alter electron behavior, at least locally in the Brillouin zone.
In graphane, the direct band gap is reduced (increased) by up to 1.2~eV (4~eV).

Our analysis shows that the origin of the band-gap modifications 
in this system is the concerted action of two geometry-induced effects.
The first cause is the hybridization between 
\textit{unoccupied} wave functions in the lowest conduction band.
This effect generally leads to lower (higher) energy state
for the bonding-type (antibonding-type) hybrid wave function,
but not always, since the graphane wavefunctions
can possess a more complex hybridization nature
than what applies in the hydrogen-like case.
The second cause is the modified electrostatic interaction 
between the hybrid wave functions and the electron density.
This cause can either increase or decrease  the energy gain (loss) 
in a bonding-type (antiboning-type) hybridization.
Moreover, it may result into a modified hybrid wave-function
where a pure bonding-type (antibonding-type) character
is lost.

Our analysis also shows that selfconsistent charge rearrangements (on the GGA level)
with respect to the ML density have no significant impact on the bandstructure.
Additional charge rearrangements described by self-consistent vdW-DF calculations\cite{Thonhauser2007}
are expected to be even smaller, justifying our use of non-selfconsistent vdW-DF.

The nature of conduction and optical absorption in BL graphane 
would be determined by the region around the $\Gamma$ point.
There, the band structure of the bilayer essentially coincides with
that of the monolayer.
Therefore, we expect graphane multilayers to behave 
electronically similar to a graphane monolayer,
at least for properties defined by a simple response.
For bulk graphane, the behavior is more complicated.
We emphasize that other vdW-bonded systems may exist
where significant band-gap modifications arise
at Brillouin-zone points having 
higher relevance for the electronic nature
of the material.

Our results for BL graphane suggest that vdW forces can have non-negligible 
\textit{indirect} effects on the overall band structure in layered or macromolecular materials.
A similar effect can be found in V2O5,\cite{LonderoRohrer}
where traditional GGA severely overestimate the $c$ lattice constant
and where vdW-DF provides a more accurate description.\cite{Londero2010}
Furthermore, we notice that vdW binding of the intrinsic semimetal graphene to
metal\cite{JacobssenGrapheneOnMetals} or semiconductor\cite{RohrerPhD, RohrerGrapheneOnSiC} surfaces 
seems to generally lead to a shift of the Fermi level,
rendering  graphene a true metal.
These observations together with the fact that 
the vdW binding strength in surface/adatom systems
(and thus presumably the strength
of the corresponding effect on  the band structure)
depends on the choice of the substrate material\cite{JacobssenGrapheneOnMetals} and on the substrate
morphology\cite{KelkkanenBenxeneOnMetals}
implies a possibility to exploit dispersive interactions also
in band-gap engineering.

Finally, a comment on the accuracy of the predicted
band-gap modifications is in order.
GGA DFT typically severely underestimates
the band gap in semiconductors or insulators.
A possible remedy is the use of the GW method
where the Kohn-Sham orbitals are used to construct 
the self-energy operator.
The GW method also contains non-local correlations
and therefore provides an alternative description
of vdW interaction.
GW calculations come at a considerably higher computational cost
than that of vdW-DF, however.
The procedure illustrated here, using vdW-DF calculations
to determine binding morphologies 
followed by band-structure calculations might well
also be adapted for GW,
pursuing  characterizations
of band-structure effects in sparse matter.

\section*{Acknowledgment}
We thank G. D. Mahan for encouragement and discussions.
Support by the Swedish National Graduate School in Materials Science (NFSM),
the Swedish Research Council (VR)
and the Swedish National Infrastructure for Computing (SNIC)
is gratefully acknowledged.




\begin{references}{}






\bibitem{BGE}
F. Capasso,
\textit{Science} \textbf{235}, 172 (1987);
F. Capasso,
\textit{Thin Solid Films} \textbf{216},  59 (1992).


\bibitem{defects}
V. H. Crespi, M. L. Cohen  and A. Rubio,
\textit{Phys. Rev. Lett.} \textbf{79}, 2093 (1997).




\bibitem{finiteSize}
B. Xu and B. C. Pan,
\textit{Phys. Rev.} B \textit{74}, 245402 (2006).



\bibitem{vdWDFreview}
D.C. Langreth, B.I. Lundqvist, S.D. Chakarova-K\"ack, V.R. Cooper, M. Dion, 
P. Hyldgaard, A. Kelkkanen, J. Kleis, Lingzhu Kong, Shen Li, P.G. Moses, 
E. Murray, A. Puzder, H. Rydberg, E. Schr\"oder, and T. Thonhauser,
\textit{J. Phys.: Condensed Matter} \textbf{21}, 084203 (2009).


\bibitem{Thonhauser2007}
T. Thonhauser, V. R. Cooper, S. Li, A. Puzder, P. Hyldgaard, and D. C. Langreth
\textit{Phys. Rev.} B \textbf{76}, 125112 (2007).

\bibitem{Londero2010}
E. Londero and E. Schr\"oder, Phys. Rev. B \textbf{82}, 054116 (2010).


\bibitem{sliding}
K Berland, T.L. Einstein, and P. Hyldgaard
\textit{Phys. Rev. B} \textbf{80}, 155431 (2009).

\bibitem{sliding2}
G. Pawin, K. L. Wong, K.-Y. Kwon, L. Bartels,
\textit{Science} \textbf{313}, 961 (2006).



\bibitem{JacobssenGrapheneOnMetals}
M. Vanin, J. J. Mortensen, A. K. Kelkkanen, J. M. Garcia-Lastra,
K. S. Thygesen, and K. W. Jacobsen, 
Phys. Rev. B \textbf{81}, 081408(R) (2010).



\bibitem{RohrerPhD}
J. Rohrer, "Formation stability and electronic structure of surfaces and int
erfaces from first principles", PhD Thesis, ISBN 978-91-7385-469-6,
Chalmers University of Technology (2010).

\bibitem{RohrerGrapheneOnSiC} 
J. Rohrer and P. Hyldgaard, unpublished.


\bibitem{KelkkanenBenxeneOnMetals}
A. Kelkkanen, B. I. Lundqvist and J. K. Norskov,
accepted for publication in PRB. 



\bibitem{vdWDF}
M. Dion, H. Rydberg, E. Schr\"oder, D. C. Langreth, and B. I. Lundqvist, 
\textit{Phys. Rev. Lett.} \textbf{92}, 246401 (2004).


\bibitem{vdWDF2}
K. Lee, \'E. D. Murray, L. Kong, B. I. Lundqvist, and D. C. Langreth,
\textit{Phys. Rev.} B 82, 081101(R) (2010).

\bibitem{C09}
V. R. Cooper,
\textit{Phys. Rev.} B \textbf{81}, 161104(R) (2010).


\bibitem{ref:Sofo_Graphane_2007}
J. O. Sofo, A. Chaudhari and G. D. Barber, 
\textit{Phys. Rev.} B \textbf{75}, 153401 (2007).


\bibitem{ref:Elias_Graphane_2009}
D. C. Elias, R. R. Nair, T. M. G. Mohiuddin, S. V. Morozov, P. Blake, M. P. Halsall, 
A. C. Ferrari, D. W. Boukhvalov, M. I. Katsnelson, A. K. Geim, and K. S. Novoselov,
\textit{Science} \textbf{323}, 5914 (2009).


\bibitem{ref:Graphene}
K. S. Novoselov, A. K. Geim, S. V. Morozov, D. Jiang, Y. Zhang, 
S. V. Dubonos, I. V. Grigorieva, and A. A. Firsov,
\textit{Science} \textbf{306}, 666 (2004).



\bibitem{BLGraphane1}
O. Leenaerts, B. Partoens, and F. M. Peeters,
\textit{Phys. Rev. B} \textbf{80}, 245422 (2009).

\bibitem{BLGraphane2}
V. I. Artyukhov and L. A. Chernozatonskii,
\textit{J. Phys. Chem. A} \textbf{114}, 5389 (2010).

\bibitem{ref:Devices1}
A. K. Geim and  K. S. Novoselov,
\textit{Nature Materials} \textbf{6}, 183 (2007).


\bibitem{ref:Devices2}
A. K. Geim,
\textit{Science} \textbf{324}, 1530 (2009).



\bibitem{BaderAnalysis1}  
G. Henkelman, A. Arnaldsson, and H. J\'{o}nsson, 
\textit{Comput. Mater. Sci.} \textbf{36}, 254 (2006)

\bibitem{BaderAnalysis2}
K. Berland, \O. Borck, and P. Hyldgaard,
submitted to \textit{Comp. Phys. Comm.}, see also arXiv:1007.3305v1;
\O. Borck, unpublished.


\bibitem{GNRExp}
M. Y. Han, B. \"Ozyilmaz, Y. Zhang, and P. Kim,
\textit{Phys. Rev. Lett.} \textbf{98}, 206805 (2007).

\bibitem{GNRModel}
V. Barone, O. Hod, and G. E. Scuseria,
\textit{Nano Lett.} \textbf{6}, 2748 (2006).


\bibitem{ChemicalModification}
Schedin et al. Nat. Mater. \textbf{6}, 652 (2007).


\bibitem{BGE_Eriksson}
M. Klintenberg, S. Leb\'egue, M. I. Katsnelson, and O. Eriksson,
\textit{Phys. Rev.} B 81, 085433 (2010).



\bibitem{GW}
L. Hedin, 
\textit{Phys. Rev.} \textbf{139}, A796 (1965);
F. Aryasetiawan and  O. Gunnarsson,
\textit{Rep. Prog. Phys.} \textbf{61}, 237 (1998).


\bibitem{ErikssonGW}
S. Lebegue, M. Klintenberg, O. Eriksson, and M.I. Katsnelson,
\textit{Phys. Rev.} B \textbf{79} 245117 (2009).



\bibitem{GraphaneQdots}
A. K. Singh, E. S. Penev and B. I. Yakobson,
\textit{ACS Nano} \textbf{4}, 3510 (2010).

\bibitem{GraphaneGrapheneJunc}
A. K. Singh and B. I. Yakobson, 
\textit{Nano Lett.} \textbf{9}, 1540 (2009).

\bibitem{hTc-SC_graphane}
G. Savini, A. C. Ferrari, and F. Giustino, 
\textit{Phys. Rev. Lett.} \textbf{105}, 037002 (2010). 



\bibitem{Pseudopotentials}
D. Vanderbilt, 
\textit{Phys. Rev. B} \textbf{41}, 7892 (1990).

\bibitem{Dacapo}
B. Hammer, O. H.  Nielsen, J. J. Mortensen, L. Bengtsson, L.B. Hansen,
A. C. E. Madsen, Y. Morikawa, T. Bligaard, A.  Christensen, and J. Rossmeisl,
available from https://wiki.fysik.dtu.dk/dacapo.



\bibitem{PBE}
J. P. Perdew, K. Burke, and M. Ernzerhof,
\textit{Phys. Rev. Lett.} \textbf{77}, 3865 (1996).




\bibitem{KpointSampling}
H. J. Monkhorst and J. D. Pack, 
\textit{Phys. Rev. B} \textbf{13}, 5188 (1976).


\bibitem{revPBE}
Y. Zhang  and W. Yang, 
\textit{Phys. Rev. Lett.} \textbf{80},  890 (1998).



\bibitem{PW86refitted}
E. D. Murray, K. Lee, and D. C. Langreth, 
\textit{J. Chem. Theory Comput.} \textbf{5}, 2754 (2009).


\bibitem{VWN}
S.H. Vosko, L. Wilk, M. Nusair, 
\textit{Can. J. Phys.} \textbf{58}, 1200 (1980).



\bibitem{Method1}
E. Ziambaras, J. Kleis, E. Schr\"oder, and P. Hyldgaard, Phys. Rev. B 76, 155425 (2007).
\bibitem{Method2}
S. D. Chakarova-K\"ack, E. Schr\"oder, B. I. Lundqvist, and D. C. Langreth, 
\textit{Phys. Rev. Lett.} \textbf{96}, 146107 (2006).
\bibitem{Method3}
K. Johnston, J. Kleis, B. I. Lundqvist, and R. M. Nieminen, 
\textit{Phys. Rev. B} \textbf{77}, 121404(R) (2008).
\bibitem{Method4}
J. Kleis, E. Schr\"oder, and P. Hyldgaard, 
\textit{Phys. Rev. B} \textbf{77}, 205422 (2008).

\bibitem{BZpoints} 
H. Ibach and H. L\"uth,
\textit{Solid-State Physics: An Introduction to Principles of Materials Science}
3rd ed. (Springer-Verlag 2002).



\bibitem{PES}
We have also mapped out the vdW-DF potential energy landscape (PES) 
that arises when one of the graphane sheets is slightly displaced 
in various directions from
this $A_+A_+$ high-symmetry configuration.
This stability test was performed for various layer separations
around the binding separation predicted by vdW-DF2.
On the PBE level, the displaced configurations essentially do not relax,
reflecting the essentially flat PBE PES at these separations.
However, adding the nonlocal vdW correction by evaluating Eq.~(\ref{eq:Ebind}),
we find that the high-symmetry configuration possesses
the lowest total energy of all considered systems.


\bibitem{LDAcomment}
We notice that full relaxation of the considered high-symmetry configurations
using LDA also predicts the $A_+A_+$ to be most favorable.
Also, the predicted LDA binding separation of $4.5$~\AA\ 
is in fair agreement with those predicted here by the various 
versions of vdW-DF
(whereas the LDA-predicted layer-binding energy of $53$ meV/cell
is considerably smaller).
However, we stress that this correspondence with vdW-DF
results coincidentally.
LDA binding in systems which (like multilayer graphane)
are bound by vdW forces 
has been assigned to unphysical
long-range exchange interactions
and LDA does not, by any means, contain vdW interactions,
see J. Harris, Phys. Rev. B \textbf{31}, 1770 (1985);
E. D. Murray, K. Lee and D. C. Langreth,
J. Chem. Theory Comput. \textbf{5}, 2754 (2009)
and references therein.







\bibitem{ref:GraphitevdW}
S. D. Chakarova-K\"ack, A. Vojvodic, J. Kleis, P. Hyldgaard, and E. Schr\"oder, 
\textit{New J. Phys.} \textbf{12} 013017 (2010).


\bibitem{ChargeTransfer}
In the present case the charge rearrangements are not
accompanied by any significant charge transfer.
Our Bader analysis \cite{BaderAnalysis1,BaderAnalysis2} 
shows that the difference in charge
before and after the charge rearrangement is $\Delta q\leq 5\cdot10^{-4}$~e/atom.
In cases where charge transfer takes place,
see for example Ref.~\onlinecite{Londero2010},
such rearrangements may have more significance.




\bibitem{LonderoRohrer}
E. Londero and J. Rohrer, unpublished.


\bibitem{PY86}
J. P. Perdew and W. Yue,
Phys. Rev. B \textbf{33}, 8800 (1986) 

\bibitem{Soler}
G. Rom\'{a}n-P\'{e}rez and J. M. Soler,
\textit{Phys. Rev. Lett.} \textbf{103}, 096102 (2009).

\bibitem{GPAW}
Selfconsistent vdW-DF is available in the opensource code 
GPAW,\cite{GPAWreview}
see \textit{https://wiki.fysik.dtu.dk/gpaw/index.html}.


\bibitem{GPAWreview}
J. Enkovaara, C. Rostgaard, J. J. Mortensen et al.
\textit{J. Phys.: Condens. Matter} \textbf{22}, 253202 (2010).






\end{references}
\end{document}